# Quantum interference without quantum mechanics

*Arend Niehaus*

Retired Professor of Physics, Utrecht University, The Netherlands

**Abstract**

A recently proposed model of the Dirac electron, which describes observed properties of the particle correctly, is in the present paper shown to be also able to explain quantum interference by classical probabilities. According to this model, the electron is not point-like, but rather an "entity" formed by a fast periodic motion of a quantum whose energy is equal to the rest energy of the electron. Only after a time span equal to the period of that periodic motion the "entity" becomes the electron, with its properties, mass, spin, charge, etc.. When in motion with respect to the observer, the "dynamic substructure" of the electron described in this way, leads to a certain time structure of its detection probability, if the space-time point of detection is taken as the space-time point of the quantum. In the typical "two slit" experimental situation, this leads to a periodic detection probability with a frequency of twice the De Broglie frequency. This result is identical to the result obtained by the quantum mechanical description of the moving electron by the free particle wave function. The different interpretations of the established interference pattern, inherent in the two alternative theoretical descriptions is outlined, and the relation between the two descriptions discussed. It is concluded that quantum interference is well explained with classical probabilities, without quantum mechanics and without paradoxes. In view of the demonstrated merits of the model on the one hand, and the new aspects regarding the established theories it implies on the other, a more thorough investigation of its role in relation to relativistic quantum mechanics, and to quantum field theory is suggested. Some of the interesting aspects are summarized.

**PACS 2010: 03.65.-w; 14.60.Cd; 02.50.Cw**

## 1. Introduction

Quantum mechanics treats the electron as a point-like particle having a certain mass and negative charge, but no structure. To describe observed phenomena correctly, it ascribes certain other properties to the particle, e.g. spin, and a wavelike nature, which are "non-classical", and cannot be "derived" [e.g., **1**]. The validity of the theory is unquestioned, however, its interpretation is still subject of debate [see, e.g., **2**], because the "quantum world" it creates contains numerous well known paradoxes.

There have been attempts to escape interpretational problems by supposing that the electron, and possibly other elementary particles also, *do* have an internal structure that possibly could explain their properties. Especially, the fact that a so called "Zitterbewegung (ZBW)" [**3**] is one of the properties arising from the relativistic quantum theoretical treatment of the free electron, has led to the proposal of a dynamic substructure [**4, 5, 6**]. Typical time scale for such a structure would be the very short period ($\tau$) of the (ZBW):

$\tau = 2\pi/\omega_{ZBW}$ with $\omega_{ZBW}=4\pi c/\lambda_c$ leading to $\tau=\lambda_c/2c$, with $\lambda_c = h/Mc$ : Compton wavelength

Theoretical analyses have indeed shown that the spin, arising in the Dirac theory, can be related to a motion, however, an *extended* (ZBW), not predicted by the Dirac theory, would be necessary to explain the properties of the electron [**6**]. Also *models* of the electron, based on the (ZBW), have been proposed [**7, 8**]. For a recent discussion we refer to [**9**].

A somewhat different approach, also based on the assumed existence of a possibly extended (ZBW), has been attempted in two recent publications [**10, 11**[1]] of the present author. The result of this approach has been the proposal of a model of an electron with a dynamic substructure. This model will be shown in the present paper to be able to explain quantum interference using classical probabilities. First, we give a short qualitative description of the model.

Starting from the assumption that spin is caused by orbital motion due to an *extended* (ZBW), a probability distribution of orientation and value of an *instantaneous orbital angular momentum* is designed, which describes spin and spin measurements in accordance with experiment. Under the assumption that the "subject" that causes the angular momenta is the quantum of a photon with momentum (Mc), probability distributions for orientation and length of the *instantaneous position vector* of the quantum are derived from the angular distributions. The instantaneous *positions* of the quantum turn out to lie on a torus around a fixed point, the torus radius ($R_t$) being equal to the radius of the circle ($R_c$) the torus axis forms around the fixed point: $R_t=R_c=(\lambda_c/4\pi)$. This special Torus is a so called Clifford Torus. The probability density of the *orientation* of the position vector is given by a cosine distribution. This is shown in **Fig.1**, which is a reproduction of Figure 3 of ref.[**11**[1]].

---

[1] Erroneously, the labels (a) and (b) of the two figures in Fig.3 were interchanged in the final typesetting.

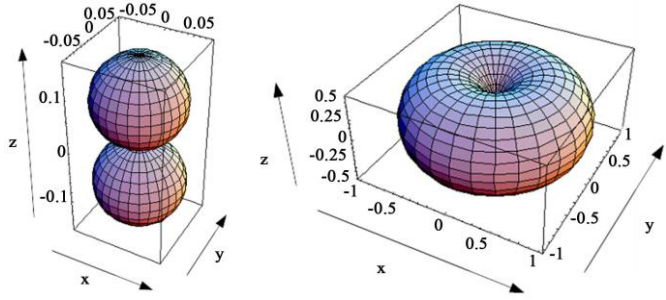

**Figure 1**: (left figure): The probability density of the orientation of the position vector is given by a cosine distribution. The probability, obtained by integration over all orientations, is normalized to unity. (right figure): The instantaneous positions of the quantum are located on a Clifford Torus. The torus radius ($R_t$) is equal to the radius of the circle ($R_c$) the torus axis forms around the fixed point: $R_t = R_c = \lambda_c/4\pi$.

The distributions are interpreted as resulting from *paths of the quantum* in space and time. These paths are the result of two combined circular motions in planes perpendicular to each other, resulting in helical path in a circle around a fixed point. The paths are closed after a characteristic period, and form an "entity" whose center is the torus center. When this entity moves with respect to the observer at different velocities, the individual paths of the quantum become different due to a velocity dependence of the two frequencies. The "entity" is identified as an individual electron, and an average of the paths of individual electrons over possible different initial conditions yields the torus surface. The properties of an individual electron are obtained by averaging the corresponding *instantaneous properties* over one period. These average properties are found to agree with properties observed for the electron. This has been verified [**11**] for the following properties: (1) Spin as the average orbital angular momentum, (2) magnetic moment, including g=2 factor, if unit negative charge is ascribed to the quantum, (3) isotropic static potential which approaches the Coulomb potential for distances large compared to $\lambda_c$, (4) momentum and kinetic energy, and of course, (5) total energy equal to $Mc^2$. The averages are exact results of elementary calculations.

The model is completely general and contains no free parameters. Its mathematical presentation is given in terms of the reduced Compton wavelength ($\lambda_c/2\pi$), and two frequencies, ($\omega_c$) and ($\omega_t$), which are determined by the rest energy of the particle described, and by the velocity of relative motion of the "entity" w.r.t. the observer. ($\omega_c$) describes the rotation of the plane containing the quantum around the Z-axis, and ($\omega_t$) describes the rotation of the quantum in this plane around the torus axis. The two frequencies are related by $\omega_c = n\, \omega_t$, and their quadratic sum is equal to the square of the (ZBW)-frequency $\omega = 4\pi c/\lambda_c$. The result of the two rotations is a helix motion of the quantum on the torus surface. When the ratio of the two frequencies is an integer, the helix describing the "entity" is closed after one period determined by the smaller of the two frequencies. An example for the closed path representing the entity in its own coordinate system for the ratio ($\omega_c/\omega_t=10$) is shown in [**Fig. 2**].

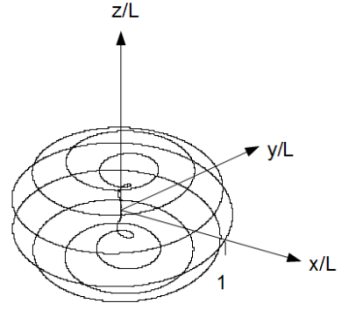

**Fig.2** : *The closed path of the quantum representing the electron for the case ($\omega_c/\omega_t=n=10$). Coordinates are given units of the reduced Compton wave length (L).*

In this paper we will demonstrate in which way the model, which represents an electron with dynamic substructure, describes observed interference phenomena of electrons correctly. As example we discuss the two slit experiment in paragraph (2). In paragraph (3) we discuss the relation between quantum theory and the model, and summarize interesting general aspects of the model.

## 2. Interference

To discuss interference, we need the representation of a free electron by the model. It is given in terms of paths in space-time of the quantum for a chosen velocity of the "entity" [11]. The velocity determines the ratio of the frequencies used in the model. This ratio has to be an integer (n). We choose the case where $\omega_c/\omega_t=n=10$, for which the paths in the moving frame of the entity is shown in **Fig.2** . The velocity of the "entity" $v=\beta c$ is the average velocity of the quantum in direction of the Z-axis, and $\beta=1/\sqrt{1+n^2}$. With the reduced Compton wave length $L=\hbar/Mc$, paths of the quantum are described by giving the coordinates of its position vector **r**(t) ={x ,y ,z}, as follows:

$\beta=1/\sqrt{(1+n^2)}$; $\omega_c=2n\beta c/L$; $\omega_t=\omega_c/n$; $\tau=4\pi/\omega_t$;                                     (1)

$x=L \cos^2(\omega_t t/2) \cos(\omega_c t)$; $y=L \cos^2(\omega_t t/2)^2 \sin(\omega_c t)$; $z=L \sin(\omega_t t)/2+\beta ct$;     (2)

A 3D-parameter representation of **r**(t) for a time span of 2.5 periods ($\tau$) is shown in the figure [**Fig.3**]. As a function of time, the path length grows in Z-direction at the speed, $v_z=\beta c+ (L/2)\omega_t \cos(\omega_t t) =\beta c(1+\cos(\omega_t t))=2\beta c \cos^2(\omega_t t/2)$.

The axis of the torus forms itself a helix around the axis formed by the center of the moving entity, also shown in **Fig3**. A point on this helix, defined by a certain value of ($\omega_c t$), moves at the speed of light in space, independent of the relative velocity (v) of the entity with respect to the observer. Therefore, one may identify this helix-path as the path of a "circulating photon" which constitutes the electron. The momentum of this photon in direction of relative motion is $Mv=\beta Mc=\beta M_0 c/\sqrt{(1-\beta^2)} =M_0 c/n$, corresponding to the de Broglie wavelength $\lambda_{DB}=h/Mv=hn/M_0 c=n\lambda_c$. The "rest energy" of the entity for (v→0), $M_0 c^2$, is the energy of the photon $\hbar\omega$. The model therefore implies the equivalence of mass and energy in a simple and convincing way.

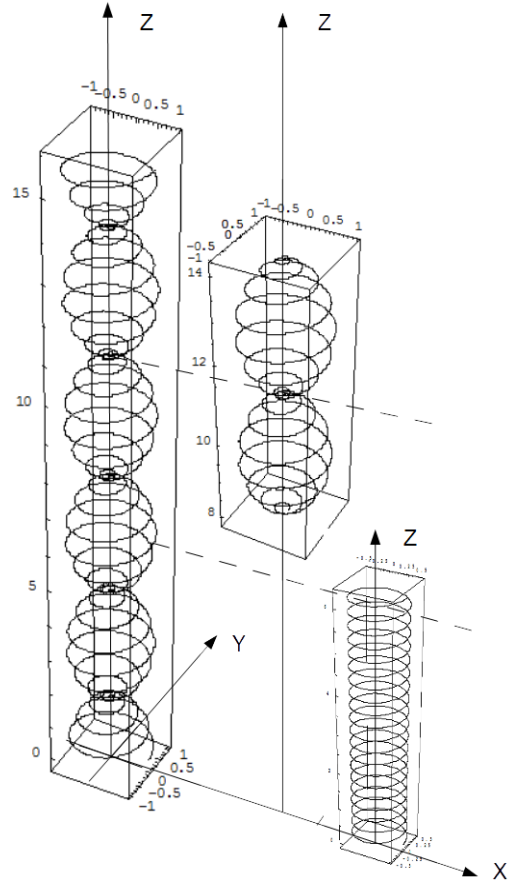

**Figure 3**: *Traces from left to right: (i), A parametric 3D plot of the positions in real space of the quantum (in units of L) during the first 2.5 periods of the (EZBW), for the case n = 10. (ii), Path of the quantum during one period, starting 1.3 periods after (t=0). (iii): Path of the "circulating photon" during the first period (see text).*
*Progress of the "entity" in z-direction is seen to proceed at the velocity v=2πL/τ=βc, while the "photon" has speed (c) on its helical path.*

For a given current in Z-direction of many "entities" starting at the same time point, there will arise a current of *quanta* through a surface positioned at a certain distance (z) from the starting point. If the diameter of the surface is not smaller than the Compton wave length, the measured current implies an integration over different initial conditions. According to the model, the current of quanta then oscillates as (normalized to one "entity per period $\tau=4\pi/\omega_t$).

$I(t) = 2 \cos^2(\omega_t/2\, t) = 2 \cos^2(\beta\, c\, t/L)$.  (3)

Since $\omega_t L/2 = \beta c = (\omega_{DB} L)$, with $\omega_{DB}$ being $2\pi$ times the De Broglie frequency, this can also be written as

$I(t) = 2 \cos^2(\omega_{DB}\, t)$.  (4)

This is also the result of the quantum mechanical treatment of the time structure of the probability to detect the particle at a certain distance (z) along its path in (Z)-direction.

Let us now consider the normal two-slit situation in the context of the model, where particles are scattered from the two slits separated by distance (B), form currents into certain directions behind the slits, and hit the surface of a screen that is oriented perpendicular to the currents and positioned at a distance large compared to the distance between the two slits [see **Fig.4**].

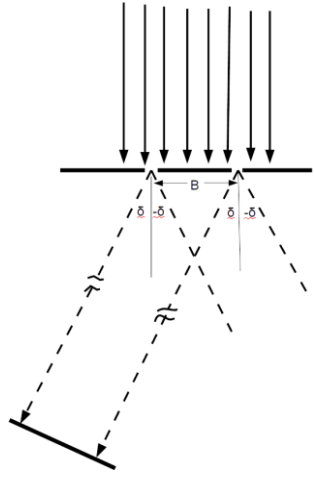

Figure 4: The two slit situation

Then, at any position on the screen determined by the scattering angle (+δ), particles from both slits will arrive. For symmetry reasons, the same current will arrive at the angle (-δ). The probability for the arrival of one particle per time unit is to be calculated from the current of combined probabilities from the two slits. Formally one could now proceed by applying the Born-rule to the current predicted by the model (see relation 4), i.e. by adding the *probability amplitude waves* having a phase difference (Dt $\omega_t/2$ ), and square the sum to obtain the total probability per unit time, W(t, Dt). The probability at position (δ), W(Dt), is then obtained by averaging over time. Assuming equal widths for the two slits, one thus obtains:

W(Dt) =2cos$^2$(Dt $\omega_t$/4)  =2cos$^2$(Dt $\omega_{DB}$/2) =2cos$^2$( β c Dt/ 2L).                    (5)

With 2πL/β = $\lambda_{DB}$ =n$\lambda_0$ , where $\lambda_{DB}$ is the De Broglie wave length, $\lambda_0$=ℏ/$M_0$c the Compton wave length, and with cDt=B sin(δ) the extra distance to be covered from one of the slits to the screen, this may be written as

W(Dt) =2cos$^2$(π c Dt/$\lambda_{DB}$) =2cos$^2$(π B sin(δ)/$\lambda_{DB}$) =2cos$^2$(π B sin(δ)/n$\lambda_0$)          (6)

This result is common to both, the quantum mechanical description, and the description by the model, because both descriptions lead to relation (4). However, obviously, the interpretation of the result is different: in case of the model, the time structure is determined by the relative phases of motion of *different quanta* coming from the two slits, while in case of the quantum mechanical description, the time structure is determined by *different phase shifts of one wave*. The "particle interference" inherent in the model, is explained by classical probabilities, as outlined below.

The Born-rule is an axiom of quantum mechanics. It implies the use of probability amplitude waves that propagate at the speed of light. In contrast, in the context of the model interference is rather understood in terms of probabilities with a defined phase that concern particles coming at a given time point from the two slits at speed (v). Since, on the average, only one particle arrives at position (δ) on the screen if two particles are scattered into direction (±δ), the probability has to be calculated for the case of two "trials".  According to classical probability theory, the total probability then is the unification of the probability currents

 I(t)= 2 a$^2$ 2cos$^2$($\omega_{DB}$  t), and I(t+Dt)=2b$^2$ 2cos$^2$($\omega_{DB}$ (t+Dt)),                    (7)

with a$^2$ and b$^2$ designating the widths of the two slits. (see relation(4)). According to the Kolmogorov axioms [**12**], the unification of these currents is generally given by the relation

W(t,Dt)= I(t)+I(t+Dt)-I(t)∩(t,Dt).                    (8)

The last term on the right hand side of (8) is the intersection of the single probabilities. We identify the intersection as:

$$I(t) \cap I(t,Dt) = (a\sqrt{2}\cos(\omega_{DB} t) - b\sqrt{2}\cos(\omega_{DB}(t+Dt)))^2. \qquad (9)$$

Relation (8) then becomes:

$$W(t,Dt) = 4a^2\cos^2(\omega_{DB} t) + 4b^2\cos^2(\omega_{DB}(t+Dt)) - 2(a\cos(\omega_{DB} t) - b\cos(\omega_{DB}(t+Dt)))^2 \qquad (10)$$

Expression (10) has to be averaged over time to obtain the probability as a function of the direction ($\delta$). The integration yields:

$$W(Dt) = (a^2 + b^2 + 2ab\cos(\omega_{DB} Dt)) \qquad (11)$$

Relation (11) is identical to the result obtained by using the Born rule, and reproduces relation (5) for the case of equal widths of the two slits. In the context of the model, relation (11) defines the probability to observe a particle at position ($\delta$), when two particles from independent "trials" pass the two slits and are scattered into directions ($\pm\delta$). The probability for detection of one particle at position ($-\delta$) is given by the same expression, and the intersection is the probability that both particles of the two trials arrive at one side, either at ($\delta$), or at ($-\delta$). In a way, therefore, the model implies a physical explanation of the Born-rule, and of the superposition principle. In addition, it clearly identifies interference as a phenomenon only emerging for *ensembles* of particles (entities), in contrast to quantum mechanics, where interference is ascribed to a "wave nature" of one particle.

In **Figure 5** we show an example using relation (6) with $B/\lambda_0 = 5n$. As a pre-factor we have inserted $\cos(\delta)$ in order to account for the geometry of the described experiment. Another dependence on ($\delta$) could occur due to a possibly $\delta$-dependent scattering probability.

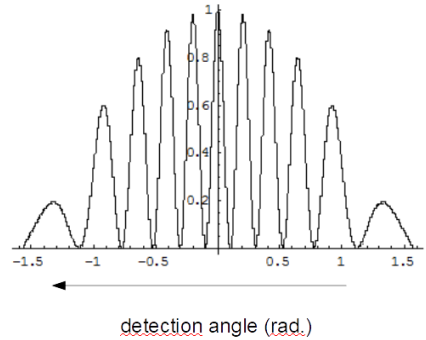

**Figure 5**: *The interference pattern calculated with the formula $W(\delta) = \cos(\delta)\cos^2(\pi\, 5\, \sin(\delta))$.*

3. Conclusion

The dynamic substructure the model ascribes to the electron explains the two slit interference, one of the main quantum phenomena, without using quantum mechanics. In a similar way, energy quantization of bound states, as well as angular momentum quantization, are predicted by the model in agreement with observation. In addition, as has been shown in the previous papers [**10, 11**], the properties of the electron, mass, spin, and magnetic moment, follow from the substructure implied by the model. Further, the paradox of "wave - particle dualism" does not arise.

In view of these successes, it is concluded that, a thorough theoretical investigation of the relation between the proposed model on one side, and quantum- and quantum-field theory on the other, would be of general interest. Such an investigation is beyond the scope of the present paper, and beyond the capacity of the author. On the other hand, we would like to point out a few interesting aspects concerning the model, not mentioned explicitly in the text.

(i) Well known paradoxes like the EPR-paradox, the wave function collapse, the wave-particle dualism, etc., do not arise. (ii) Since the model does not contain any free parameter, it seems possible that, the dynamic structure at ultrashort times it implies, is the *general cause* for the observed quantum behavior of matter.(iii) The model describes "mass" as a quantity resulting from a certain type of autonomous motion of a quantum, similarly as it treats "light" as an entity resulting from another type of autonomous motion of a quantum, whereby "mass" is characterized by two frequencies of periodic motion, while "light" is characterized by one frequency of periodic motion. This suggests that, in general, "entities" like elementary particles and photons, may be the consequence of closed paths with different topologies of a quantum, and "emerge" as averages over short periods of autonomous motion. (iv) The common origin of mass and light suggested by the model, may also be interesting in connection with attempts to link quantum- and quantum-field theory to general relativity. For a recent discussion we refer to [**2**].

## References

[1] MESSIAH A.in "Quantum Mechanics" Vol.II (North Holland Publ. Comp. 1964) pp.540
[2] KHRENNIKOV A.: Found Phys **47**:1077–1099 (2017)
[3] SCHROEDINGER E.: *Sitzungber. Preuss. Akad. Wiss. Phys.-Math. Kl.* **24**, 418 (1930
[4] HESTENES D.: Am. J. Phys. **47**,399–415 (1979).
[5] HESTENES D.: Found. Phys. **20**, 1213-1232 (1990)
[6] HESTENES D.: Annales de la Fondation Louis de Broglie, **28,** 390-408 (2003)
[7] BARUT A.O., SANGHI N.: Phys. Rev. Lett. **52**, 2009 - 2012 (1984)
[8] VAZ JR, J.: Phys. Lett. **B344** (1-4), 149-157(1995)
[9] ] PAVSIC M. , RECAMI E. , WALDYR A., RODRIGES JR G. , MACCARRONE D., RACCITI F. , SALESI G.: Phys.Lett. **B318** , 481-488 (1993)
[10] NIEHAUS A.: Found. Phys. **46**: 3-13 (2016)
[11] NIEHAUS A.: Journal of Modern Physics **8**, 511-521 (2017)
[12] BAUER H.: in „*Wahrscheinlichkeitstheorie*" (de Gruyter, Berlin, New York 2002)